\documentclass[]{spie}  
\usepackage{xcolor}

\usepackage{amsmath,amsfonts,amssymb}
\usepackage{graphicx}
\usepackage[colorlinks=true, allcolors=blue]{hyperref}
\usepackage{wrapfig}
\usepackage{float}
\usepackage{placeins}
\usepackage{multirow}
\title{RF Signal Transformation and Classification using Deep Neural Networks}

\author{Umar Khalid}
\author{Nazmul Karim}
\author{Nazanin Rahnavard}
\affil{Department of Electrical and Computer Engineering, University of Central Florida, Orlando, USA}

\authorinfo{Further author information: (Send correspondence to Umar Khalid)\\Umar Khalid.: E-mail: umarkhalid@knights.ucf.edu\\
Code and RF1024 Dataset are available at \url{ https://github.com/UmarKhalidcs/RF\_Classification}}

\begin{document} 
\maketitle

\begin{abstract}
Deep neural networks (DNNs) designed for computer vision and natural language processing tasks cannot be directly applied to the radio frequency (RF) datasets. To address this challenge, we propose to convert the raw RF data to data types that are suitable for off-the-shelf DNNs by introducing a convolutional transform technique. In addition, we propose a simple 5-layer convolutional neural network architecture (CONV-5) that can operate with raw RF I/Q data without any transformation. Further, we put forward an RF dataset, referred to as RF1024, to facilitate the future RF research. RF1024 consists of 8 different RF modulation classes with each class having 1000/200 training/test samples. Each sample of the RF1024 dataset contains 1024 complex I/Q values. Lastly, the experiments are performed on the RadioML2016 and RF1024 datasets to demonstrate the improved classification performance.
\end{abstract}


\section{INTRODUCTION}
\vspace{3mm}
\label{sec:intro} 
\indent Traditionally, modulation recognition and radio signal classification have been accomplished by specialized signal processing based feature extractors, which employ either statistical-learned or analytically-derived decision boundaries\cite{headley2010asynchronous,dobre2007survey}. In recent years, however,  deep learning has emerged as a robust feature extraction technique with wide-ranging applications from computer vision to natural language processing and has replaced the previously popular feature extraction methods such as scale-invariant feature transform (SIFT)~\cite{hertel2015deep, kamper2015unsupervised,iqbal2014dual}.\\
\indent Researchers have also studied the feasibility of the deep neural networks (DNNs) in the realm of modulation recognition \cite{zhou2020deep, li2018robust,west2017deep,o2016convolutional, zeng2019spectrum}. It has been reported that  deep learning models can be applied to the radio frequency (RF) modulation classification task similar to their applications for image or action classification. It has been further studied that DNNs are more accurate and sensitive when used as signal identifiers as compared to the established conventional signal processing approaches.  Therefore, DNNs, especially convolutional neural networks (CNNs), are being widely explored for the 
RF modulation classification, which is a critical task in the field of wireless communication. \\
\indent However, CNNs are generally proposed for image and computer vision tasks. For example, CNNs designed for ImageNet \cite{deng2009imagenet} or CIFAR \cite{krizhevsky2009learning} datasets cannot be directly applied to the RF dataset. To address this issue, we propose a \emph{convolutional transform} method that converts the I/Q signal into an output of size $ 2 \times W \times W$, where $W \in \{128,256,1024\}$, depending on the data used for experimentation. This paper further investigates the Short-Time Fourier Transform (STFT) \cite{nawab1983signal,durak2003short} to obtain the time-frequency image of the RF signal of size $2 \times W \times W $, where $W$  $\in$  $\{28,224\}$. Another contribution of this work is the collection of a real dataset, referred to as RF1024, for 8 RF modulation classess. Lastly, this work also proposes a simple 5-layer CNN (CONV-5) architecture for the I/Q data input  by substantially reducing the number of parameters as compared to the more deeper and sophisticated models. To compare the effectiveness of the proposed method, experiments are performed on the RF1024 dataset and the publicly available synthetic RadioML2016 dataset\cite{rml_datasets}. \\
\indent In our experiments, we have shown that by using the proposed transformations, the RF modulation recognition performance can be significantly improved. We show that the classification accuracy for the RF1024 dataset is nearly 99 $\%$ provided that SOTA CNN models such as ResNet18 \cite{he2016identity} and VGG16\cite{Simonyan15} are used.\\
\indent Furthermore, the performance of the proposed 5-layer CNN (CONV-5) architecture for the raw I/Q input is shown to be comparable to the other proposed models in the literature. To compare the effectiveness of the proposed method, experiments are performed on both RF1024 and RadioML2016 datasets.  \\
\indent The  rest  of  the  paper  is  organized  as  follows.  Section \ref{relatedwork} discusses the related work. Section \ref{approach} presents the proposed deep learning architecture and transformation techniques. Experimental results are presented in Section \ref{experiments} and Section \ref{conclusion} concludes the paper.  \\
\section{RELATED  WORK}
\vspace{3mm}
\label{relatedwork}
 As the security and safety of wireless communication systems have gained paramount importance in recent years~\cite{lin2020threats,shi2019deep,shi2019generative}, researchers have started exploring DNNs to design robust RF recognition systems \cite{o2016convolutional, west2017deep,zeng2019spectrum, li2018robust}. Traditionally, RF identification methods are mainly based on RF fingerprint extraction and signal sampling \cite{xu2010likelihood, grimaldi2007automatic}. Such traditional methods first employ a feature extraction method. Next, those extracted features are used as standard impressions to identify the type of the signal\cite{headley2010asynchronous, dobre2007survey}. However, recent studies have suggested that traditional modulation recognition methods are not robust and tend to fail in low signal-to-noise (SNR)  scenarios \cite{shi2019deep}. 

Moreover, such conventional techniques require preset computational formulation. Provided that such methods are based on preset specific formulas, the recognition is biased based on the pre-defined strategy. Machine learning has addressed the shortcomings of such orthodox methods and has found wide ranging applications in the domain of wireless communication~\cite{kim2020over,shi2019deep}. Inspired by their success in computer vision and natural language procssing (NLP)~\cite{9596233,zafar2019detect}, DNNs are recently being extensively studied in the realm of RF signal processing. 


Contrary to traditional RF recognition systems, DNN-based modulation recognition models have proven to improve the RF signal identification in dynamic spectrum environment. \cite{shi2019deep,o2016convolutional,west2017deep,zhou2020deep}. \\
\indent Therefore, more recently, DNNs have also been employed in complex spectrum settings by applying  the sophisticated computer vision techniques on the RF data. Authors in~\cite{patel2020data} studied the spectrum data augmentation by generative adversarial networks (GANs). ~\cite{vashist2012simulation,seyman2013channel} explored the channel estimation using feed forward neural networks (FNNs) and ~\cite{shi2019deep} studied the RF signal classification in dynamic spectrum environments using CNNs. RF modulation classification and  RF fingerprint recognition have been recently explored ~\cite{zong2020rf, tang2021radio} as well. The goal of modulation classification is to identify and classify a given RF signal to a known modulation type whereas RF fingerprint recognition is a method to identify the annunciator or the transmitting circuit. Authors in ~\cite{kim2020over,xu2020deep} further evaluated the  performance of  modulation  classification  with  over-the-air measurements of RF signals.  Moreover, ~\cite{shi2019deep} demonstrates how the performance in real-world systems varies by introducing a variety of interference in the real-world RF settings. Inspired by these studies, we have proposed a CNN-based RF signal classifier that incorporates raw I/Q signal input. We have further introduced a convolutional transform approach to modify the input in a form that could help train the SOTA DNN models such as ResNet~\cite{he2016identity} and VGG-16~\cite{Simonyan15}. \\

\section{Convolutional Transform and CNN Architecture}
\begin{figure*}[t]
\centering
\vspace{3mm}
\includegraphics[height=4.7cm, width=17cm,trim={0cm 3cm 1.5cm 9cm}, clip]{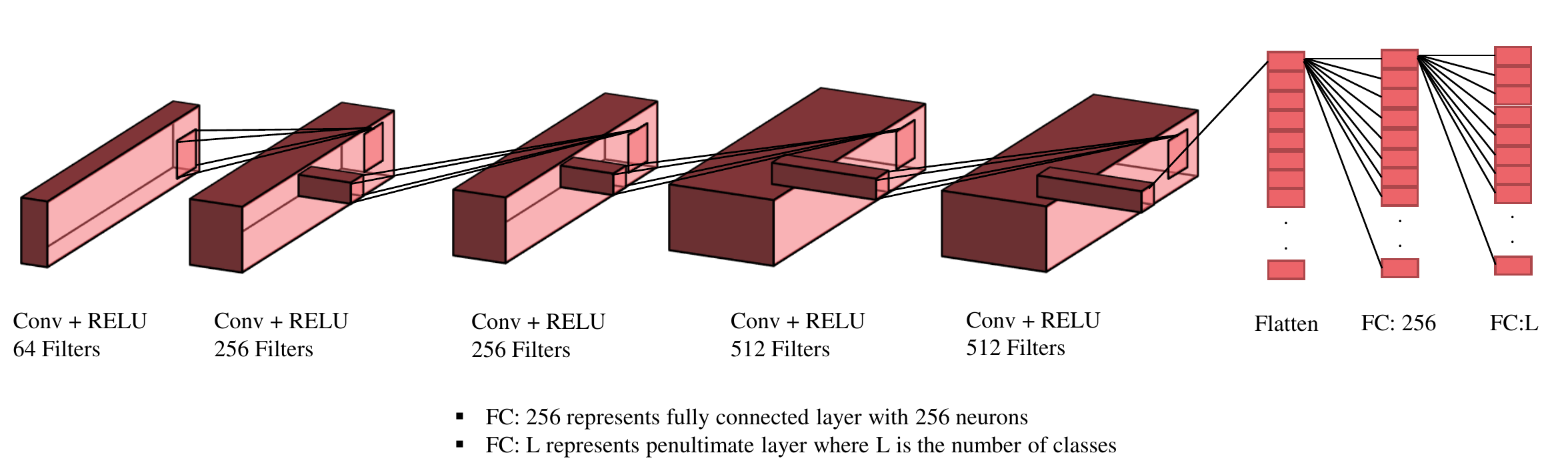}
\caption{Proposed CONV-5 architecture. Here, Kernel Size =3, Stride=1 and Padding=1 for all 5 convolution layers. }
\label{fig1}
\end{figure*}

\vspace{3mm}
\label{approach}
Suppose that signals are transmitted from a transmitter and our aim is to determine the modulation type of the transmitted signals using a receiver with $N_r$ antennas. If $\mathbf X$ is the transmitted signal, the received baseband signal $\mathbf S$ is given as:
\begin{equation}
    \mathbf {S= ZX+n},
\end{equation}
where $\mathbf{Z}$ is a complex vector with dimension $N_r \times 1$ and $\mathbf{n}$ denotes the Gaussian noise. In one observation interval, each antenna yields $N$ signal samples. Hence, $1 \times N$ vector is formed with complex values that can be decomposed into matrix $M$ with dimensions $2 \times N$. The first row of $M$ corresponds to the in-phase ($\mathbf I$) component, while the second row of $M$ contains the quadrature ($\mathbf Q$) component. The matrix $M$, and its corresponding label which identifies the modulation type forms one training sample.\\
\indent Next, we will formulate the optimization problem for supervised training of a DNN network. The supervised optimization of the parameters $\theta$ of a neural network can be achieved by reducing the cost function $J(\theta)$ using the training samples. The cost function for CNN training can be written as:
\begin{equation}
J(\theta) = \mathbb{E}_{(x_i,y_i)} L(f(x_i;\theta),y_i))),
\end{equation}
where $x_i$ is the $i^{th}$ training sample and $y_i$ is the corresponding label of the $i^{th}$ training sample. $f$ represents the DNN, $L$ is the cross-entropy loss,  and $\mathbb{E}$ is the expectation over the training data distribution. \\
\indent Next, we propose a simpler CNN architecture (CONV-5) as shown in Fig.~\ref{fig1}, which is custom-made for RF data with raw I/Q values. The proposed model has been carefully designed to take I/Q data as input, as opposed to vanilla ResNet or VGG models that cannot support such raw RF data. This is because the dimensions of the RF signals is not the same as those of the images. As compared to other studies in RF signal classification, our model is much simpler with less number of parameters and gives comparable performance. The CONV-5 shown in Fig.~\ref{fig1} can be trained similar to any other CNN architecture using cross-entropy loss function for signal classification. 




Alternatively, to enhance the classification performance by using well-known off-the-shelf CNN models, mainly designed for computer vision tasks, we need to first apply some transformation on the raw I/Q data before feeding it to the CNN. We have proposed two transformation techniques that can be used to convert I/Q signal dimensions such that the dimensions of the transformed data matches that of images. \\
\indent First, the steps involved in the convolutional transform (CT) are listed below:
\begin{itemize}
  \setlength{\itemsep}{1pt}
  \setlength{\parskip}{0pt}
  \setlength{\parsep}{0pt}
    \item Input shape: (1,2,1024)
    \item  Convolutional layer with 256 filters with kernel size of (3,3), padding=1 and stride =1
    \item   Axis swipe 1$\rightleftharpoons$2
    \item 2D MaxPooling layer with size (1,4)
    
\end{itemize}

\indent Dimensions used above are based on RF1024 dataset, where the input size is 2 $\times$ 1024 and the transformed output generated will be of size 2 $\times$ 256 $\times$ 256. In the case of RadioML2016 dataset, the input dimension is 2 $\times$ 128. So, the convolutional layer in the convolutional transform above will have 32 filters and the final transformed output will be 2 $\times$ 32 $\times$ 32, similar to the CIFAR dataset in computer vision.\\
\indent Second, we have also explored the signal-processing-based transformation technique which is short-time Fourier transform (STFT). STFT can also help to fully expose the time-frequency properties of the RF signal. To implement STFT, the given RF signal is converted from a time-domain format to the time-frequency format to obtain the signal spectrum of the given RF sample. STFT not only reflects the features of the signal in the frequency domain, but also highlights the frequency domain variations over time that can help the CNN to do better feature learning in the supervised setting. STFT of a given signal is given as: 

\begin{equation}
    \texttt{STFT}\{x(t)\} (\tau, f ) =X(\tau, f) = \int_{-\infty}^{\infty} x(t) w(t-\tau) e^{-j2\pi f t} dt 
\end{equation}
\indent where $w (\tau)$ is the window function which in our study is Hann window \cite{}, $x(t)$ is the signal to be transformed, and $f$ represents the frequency. For RF1024, using the window size of 128 and overlap of 112, STFT module gives a transformed output of size  2 $\times$ 256 $\times$ 256.\\
\begin{wrapfigure}{R}{0.5\textwidth}
  \begin{center}
  \vspace{-10mm}
    \includegraphics[width=0.4\textwidth, trim={5cm 1cm 7cm 2cm},clip]{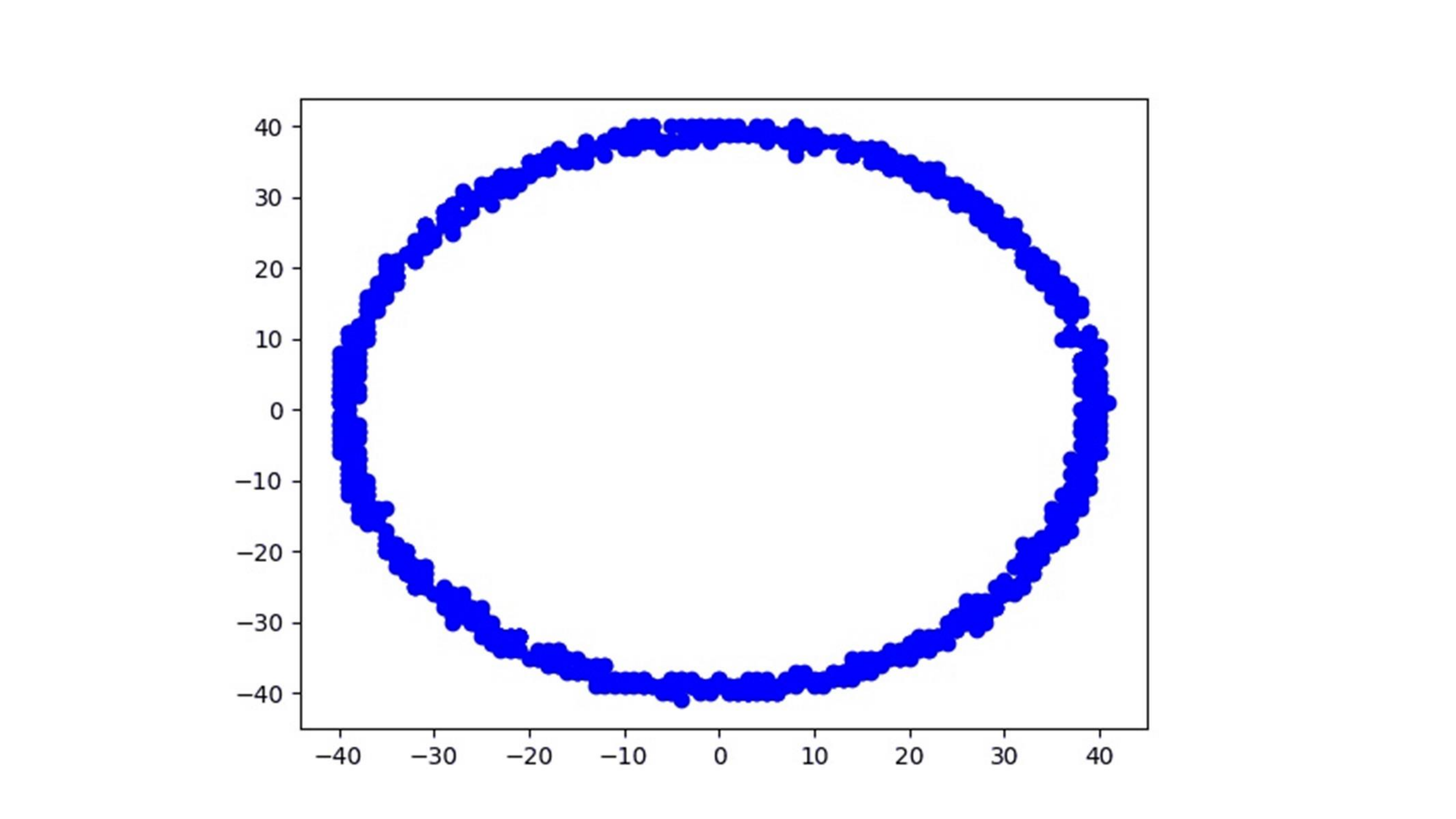}
  \end{center}
  \vspace{-5mm}
  \caption{The constellations of raw I/Q values of the FM signal sample.}
\label{RF1024}
 \vspace{-9mm}
\end{wrapfigure}

\indent As mentioned in the above section, one of the contributions of this work is the preparation of RF1024 dataset comprising of 8 modulation classes including QAM, FSK, FM, GSMK, OFDM, 4-FSK, 4-PSK and QPSK. An I/Q constellation diagram belonging to the modulation class FM is shown in Fig.~\ref{RF1024}. Next, we will evaluate the performance of the proposed architecture and transformation techniques on the prepared dataset, RF1024, and publicly available dataset, RadioML2016.

\section{Experiments}
\label{experiments}
In this section, we evaluate our proposed method through extensive experimentation on different datasets and multiple architectures. Next, we will discuss the datasets, CNN architectures and experimental details in details. 

\subsection{Datasets}To conduct our experiments, we use RF1024 amd RadioML2016 datasets \cite{rml_datasets}. In RadioML dataset, each sample consists of 128 complex valued data points, while in RF1024, each sample contains 1024 complex valued data  points. Each data point has a real and an imaginary component. RadioML dataset conatains 11 modulation classes which are: QPSK, 8PSK, QAM16, QAM64, CPFSK, GFSK, PAM4, WBFM, AM-SSB, BPSK and AM-DSB. To prepare RadioML2016 dataset, the data has been collected over a wide range of SNRs ranging from -20 dB to 18 dB. The introduced RF1024 consists of 8 modulations categories listed as: QAM, FSK, FM, GSMK, OFDM, 4-FSK, 4-PSK and QPSK with each class having 1000/200 training/test samples. Each sample of the RF1024 dataset contains 1024 complex I/Q values. 
\begin{table*}[h!]
	\caption{CNN classifier accuracy (averaged over all signal types).Here, C5 represent CONV-5, RN18-ST represents ResNet-18 with STFT and RN18-CT represents ResNet-18 with convolutional transform.}
	\label{tab:table1}
	\centering
	{\small
		\begin{tabular}{||c|c||c|c||}
		\hline
			\multirow{2}{*}{SNR (dB)} & Accuracy (\%) & \multirow{2}{*}{SNR (dB)} & Accuracy(\%) \\
			 \cline{2-2} \cline{4-4}
			 & VT \cite{o2016convolutional}/CLD \cite{west2017deep}/C5/RN18-ST/RN18-CT &  & VT2 \cite{o2016convolutional}/CLD\cite{west2017deep}/C5/RN18-ST/RN-18-CT \\
			 \hline
			-20 & 9.5/9.6/10.3/{\bf{11.2}}/10.5 & 0 &
			70.4/81.2/82.5/83.4/\bf{88.4} \\ 
			-18 & 9.1/10.8/11.1/{\bf{11.3}}/10.3 & 2 & 70.8/83.9/85.0/84.2/\bf{89.6} \\
			-16 & 10.0/10.6/11.9/{\bf{12.3}}/11.2 & 4 & 72.2/84.2/85.8/84.9/\bf{89.4} \\
			-14 & 10.4/14.0/15.1/{\bf{15.5}}/14.9 & 6 & 72.7/84.5/86.3/84.8/\bf{90.0} \\
			-12 & 15.0/21.5/18.2/{\bf{22.2}}/19.8 & 8 & 71.7/84.7/85.9/85.1/\bf{89.7} \\
			-10 & 22.7/31.5/28.8/{\bf{32.5}}/28.1 & 10 & 73.2/85.0/87.1/85.2/\bf{90.6} \\ 
			-8 & 34.8/45.3/43.5/{\bf{47.0}}/42.4 & 12 & 72.3/85.0/85.5/85.5/\bf{90.9} \\
			-6 & 49.3/63.4/58.3/{\bf{63.5}}/62.1 & 14 & 72.9/85.1/84.7/85.5/\bf{90.2} \\
			-4 & 58.9/71.2/72.1/72.5/\bf{75.7} & 16 & 72.5/85.2/84.8/85.5/\bf{90.5} \\
			-2 & 64.9/78.1/78.9/81.3/\bf{84.7} & 18 & 72.3/85.2/84.8/85.6/\bf{90.2} \\ \hline
		\end{tabular}
	}
\end{table*}

\begin{figure*}[t]
\centering
\vspace{3mm}
\includegraphics[height=8cm, width=16cm,trim={0cm 1cm 0cm 0cm}, clip]{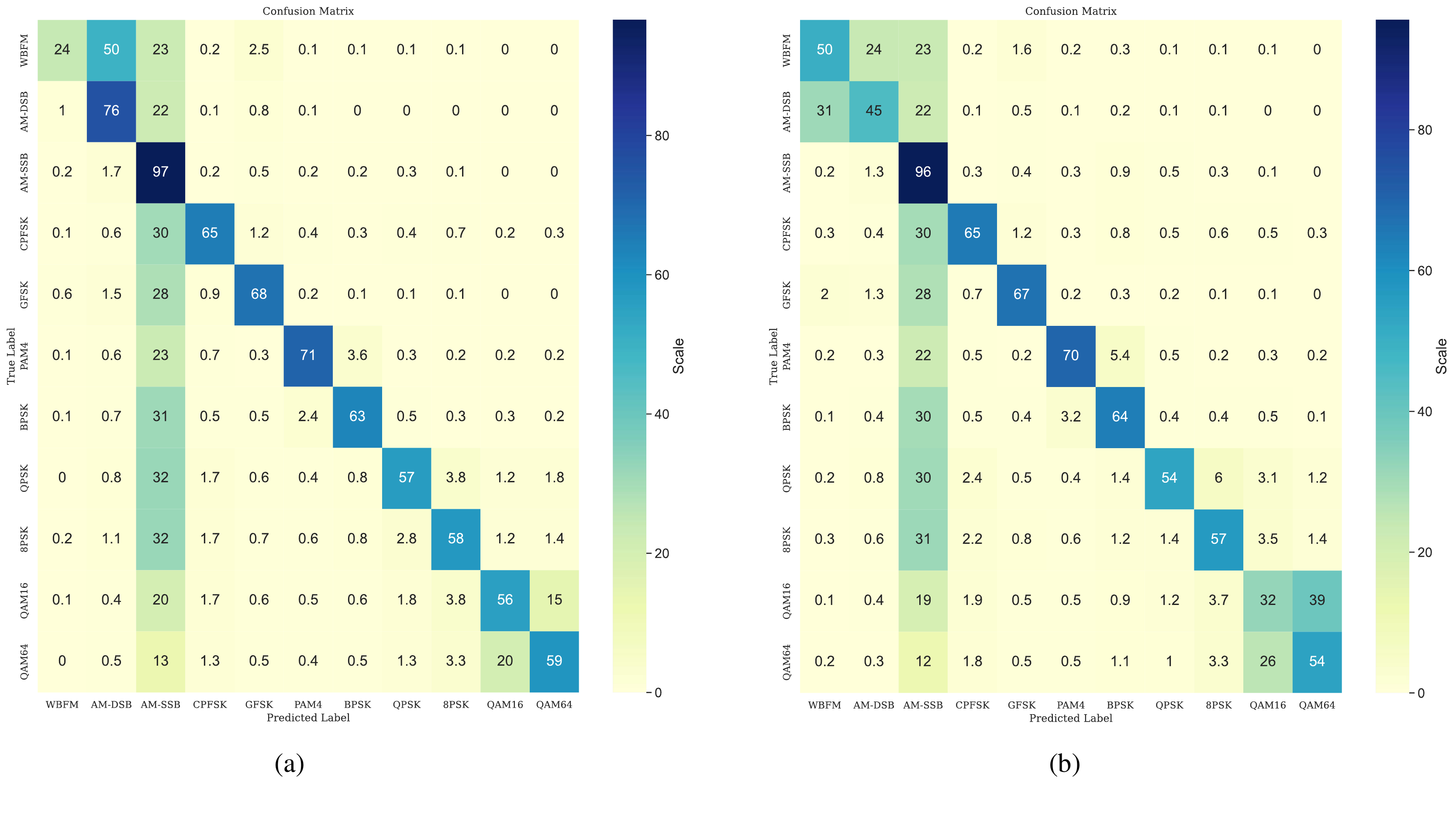}
\caption{Confusion matrices on test set of RadioML2016 dataset. All values are rounded off to display two digits. {\color{red}{(a)}} Confusion matrix obtained by evaluating ResNet-18 on RadioML-2016 test dataset. {\color{red}{(b)}} Confusion matrix obtained by evaluating CONV-5 on RadioML2016 test dataset. }
\label{tab:fig1}
\end{figure*}
\begin{figure*}[]
\centering
\includegraphics[height=8cm, width=16cm,trim={0cm 1cm 0.5cm 0cm}, clip]{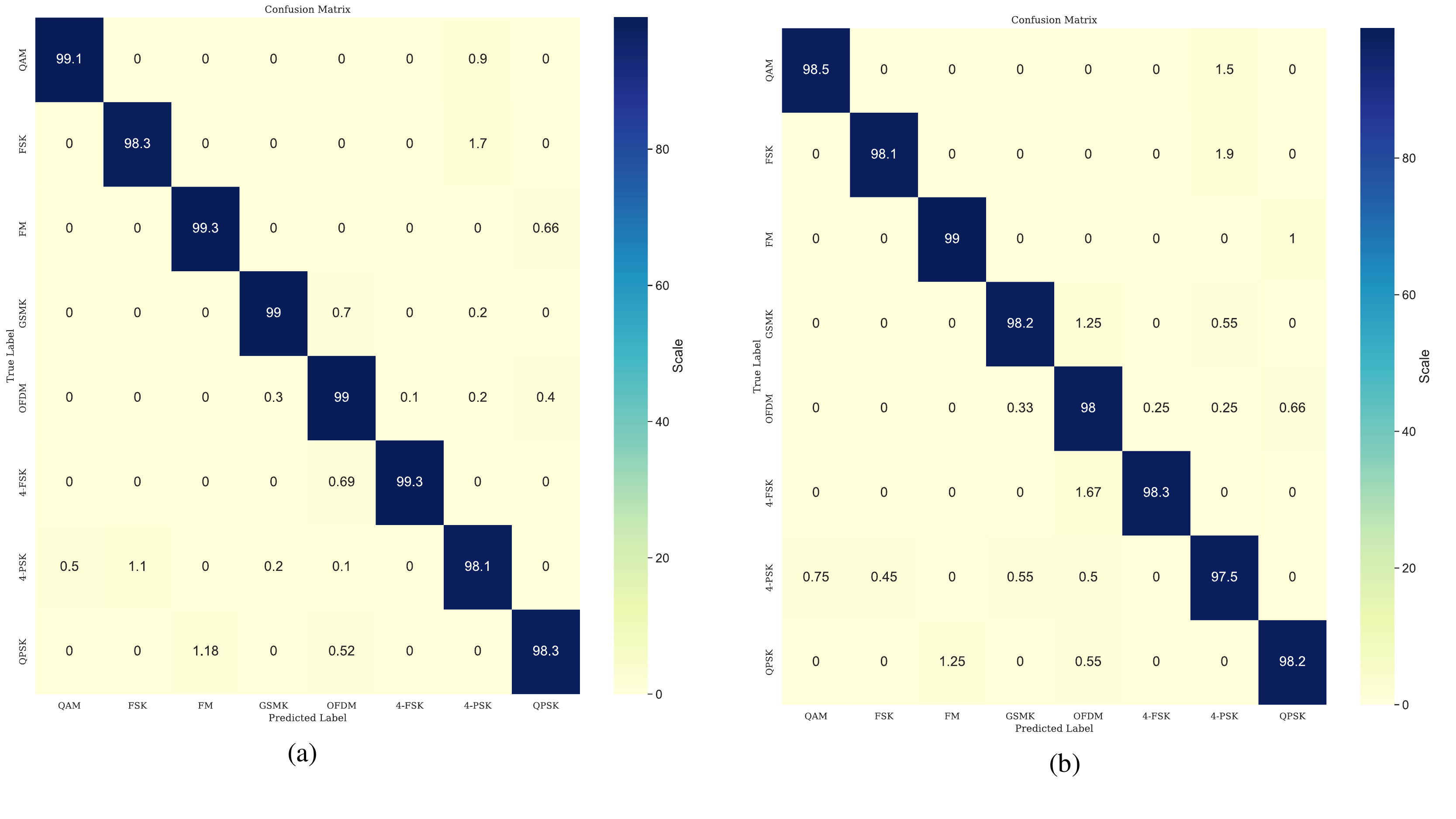}
\caption{Confusion matrices on test set of RF1024 dataset. All values are rounded off to display three digits. {\color{red}{(a)}} Confusion matrix obtained by evaluating ResNet-18 on RF1024 test dataset. {\color{red}{(b)}} Confusion matrix obtained by evaluating CONV-5 on RF1024 test dataset. }
\label{tab:fig2}
\end{figure*}
\subsection{Architecture}
The results reported are based on the proposed CONV-5 architecture. Moreover, ResNet-18 is used to evaluate performance with the proposed transformation techniques. 
\subsection{Results}
To evaluate the proposed architecture and transformations, we train the model for 10 epochs with learning rate 0.1 in all our experiments and the model with the minimum loss on the test set is then used for inference.\\
\indent First, CONV-5 is evaluated on RadioML2016. The comparison with other proposed modulation classification architectures has been provided in Table ~\ref{tab:table1}. CONV-5 achieves similar performance as compared to the methods ~\cite{o2016convolutional,west2017deep} designed for raw I/Q input with less number of model parameters. Trainable parameters of Conv-5 are 5067019. Moreover, the results of ResNet-18 with STFT and convolutional transform has also been shown in Table ~\ref{tab:table1}. In case of RadioML2016 dataset, the performance of CT is better in high SNR settings while STFT has slightly better performance in low SNR scenarios.  \\
\indent The confusion matrix for ResNet-18 while testing on the test set by incorporating all 11 classes of the RadioML2016 and 8 classes of RF1024 dataset has been shown in Fig.~\ref{tab:fig1}. In addition, the confusion matrix generated by evaluating CONV-5 on test set has been shown in Fig.~\ref{tab:fig2}.

\section{Conclusion}
\vspace{3mm}
\label{conclusion}
This paper investigates the utility of transformations such as convolutional transform and short-time Fourier transform on the RF data. This work establishes that convolutional transform is an effective technique to utilize the architectural design of SOTA vanilla CNN model on RF modulation recognition task. ResNet-18 with convolutional transform outperforms the CNN based architectures proposed in earlier study for RF signal identification. Hence, this work established the off-the-shelf CNN models yield better modulation recognition performance provided that RF data is appropriately transformed.  
\section{Acknowledgments}
This work is supported by Leonardo DRS.\\
\vspace{-7mm}
\bibliography{report} 
\bibliographystyle{spiebib} 

\end{document}